\documentclass[aps,prb,twocolumn,showpacs]{revtex4}
\usepackage[latin1]{inputenc}
\usepackage{graphicx}
\usepackage[pdftex,hypertexnames=false,pdfpagemode=UseThumbs,										% UseOutlines, UseThumbs, FullScreen, UseNone
linkbordercolor={1 1 1}, citebordercolor={1 1 1},pdfstartview=FitH]{hyperref}
\usepackage{amsmath}
\usepackage{bbold}
\DeclareMathAlphabet{\mathpzc}{OT1}{pzc}{m}{it}
\renewcommand{\vec}[1]{\boldsymbol{#1}}

\begin{document}

\title{Acoustic phonons and spin relaxation in graphene nanoribbons}
\author{Matthias Droth}
%\email{matthias.droth@uni-konstanz.de}
\author{Guido Burkard}
%\email{guido.burkard@uni-konstanz.de}
\affiliation{Department of Physics, University of Konstanz, 78457 Konstanz, Germany}
\begin{abstract}
Phonons are responsible for limiting both the electron mobility and the spin relaxation time in solids and provide a mechanism for thermal transport. In view of a possible transistor function as well as spintronics applications in graphene nanoribbons, we present a theoretical study of acoustic phonons in these nanostructures. Using a two-dimensional continuum model which takes into account the monatomic thickness of graphene, we derive Hermitian wave equations and infer phonon creation and annihilation operators. We elaborate on two types of boundary configuration, which we believe can be realized in experiment: (i) fixed and (ii) free boundaries. The former leads to a gapped phonon dispersion relation, which is beneficial for high electron mobilites and long spin lifetimes. The latter exhibits an ungapped dispersion and a finite sound velocity of out-of-plane modes at the center of the Brillouin zone. In the limit of negligible boundary effects, bulk-like behavior is restored. We also discuss the deformation potential, which in some cases gives the dominant contribution to the spin relaxation rate $T_1^{-1}$.
\end{abstract}
\maketitle

\section{Introduction}
Its interesting electronic, mechanical, and thermal properties have made graphene a promising candidate for a wide range of applications, including ballistic transistors as well as spintronics and nanoelectromechanical devices and heat management.\cite{Novoselov2004,Lin2010,Lemme2007,Trauzettel2007,GarciaSanchez2008,Balandin2008,Nika2009} There are, however, a number of challenges: (i) for epitaxial graphene, which is desirable for a controlled, large-scale production, the strong coupling to a substrate compromises these properties, (ii) graphene has no band gap, a handicap for typical semiconductor applications, and (iii) acoustic phonons limit the carrier mobility relevant for transistor functions.\cite{Ouyang2008,Finkenstadt2007,Farmer2009,Betti2011,Yoon2011}

The first issue can be overcome by removing substrate material from underneath the carbon layer such that a trench is formed and the electronic properties of free-standing graphene are restored.\cite{Meyer2007,Nair2008,Bolotin2008,Shivaraman2009,Lima2010} The second challenge is met by graphene nanoribbons (GNRs), graphene strips with a width at the nanometer scale (e.g., $L\sim1\,\mu {\rm m}$, $W\sim30\,{\rm nm}$) which can exhibit a band gap.\cite{Wang2011,Han2007} Combining these advantages, the free-standing GNR obtained from epitaxial graphene on a trenched substrate is a very interesting design that deserves a detailed discussion of its phonons.% However, it is known that phonons limit the carrier mobility and are thus relevant for transistor functions \cite{Ouyang2008,Finkenstadt2007,Farmer2009}.

\begin{figure}[t]\centering\includegraphics[width = 0.46\textwidth]{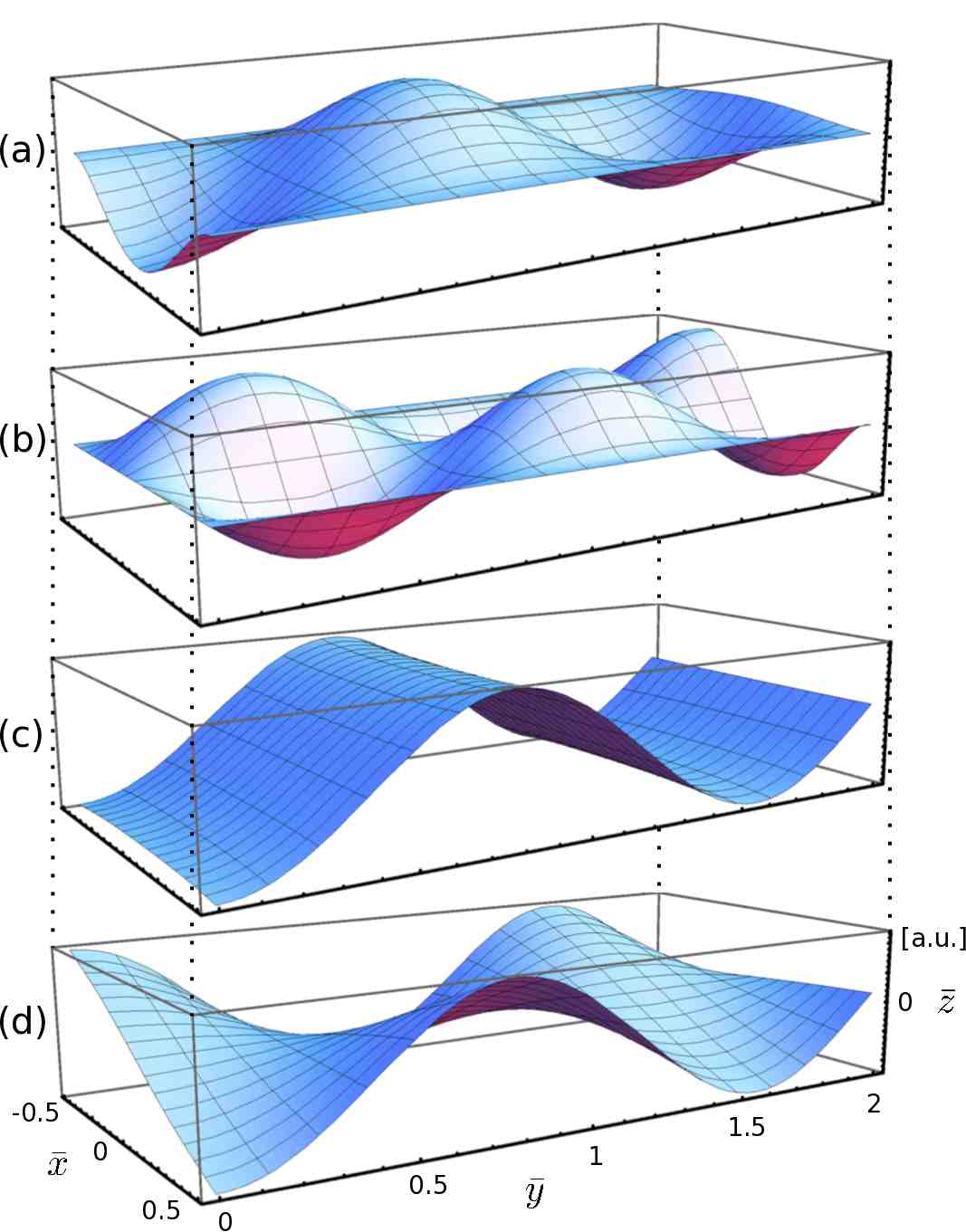}\caption{(Color online) Displacement field $u_z$ of out-of-plane modes. The dimensionless coordinates are $\bar{x}=x/W$, $\bar{y}=y/W$, and $\bar{z}=z/W$. (a),(b) Fixed boundaries. (c),(d) Free boundaries. (a),(c) Fundamental mode. (b),(d) First overtone.}\label{pic5}\end{figure}

\begin{figure}[t]\centering\includegraphics[width = 0.48\textwidth]{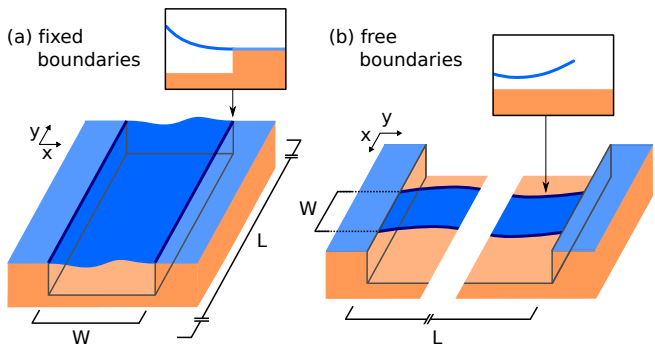}\caption{(Color online) Two nanoribbon configurations where graphene (blue) is spanned over a trenched substrate (orange): (a) fixed and (b) free boundaries. The coordinate system is chosen in such a way that the undeformed ribbon lies in the $xy$ plane and that the $y$ and ribbon axes coincide. We assume the ribbon length to be much larger than the ribbon width ($L\gg W$) and parallel lateral boundaries (dark blue) at $x=\pm W/2$.}\label{pic1}\end{figure}

In this paper, we use a continuum model to study the acoustic phonon properties 
and displacement fields $\vec{u}(\vec{r})=(u_x,u_y,u_z)$ (out-of-plane modes shown in Fig. \ref{pic5})
of two different types of GNR that we think can be realized in experiment: (i) extended graphene that covers a thin trench, resulting in a GNR parallel to the trench and with fixed lateral boundaries, Fig. \ref{pic1} (a); (ii) a strip of graphene that stretches over a wide trench, leading to a GNR perpendicular to the trench and with free lateral boundaries, Fig. \ref{pic1} (b). For both setups, we derive the low-energy acoustic phonon spectra from a continuum model that respects the monatomic structure of graphene and write down the quantum mechanical form of these phonons.

Our results can be probed experimentally via established techniques like electron energy loss spectroscopy or Brillouin light scattering.\cite{Oshima1988,Mohr2007}
In addition to the electron mobility, the phononic behavior is essential for carbon-based nanoelectromechanical systems.\cite{GarciaSanchez2008,Steele2009} A recent example where the electron-phonon coupling has been observed experimentally is the Franck-Condon blockade in suspended carbon nanotube 
quantum dots.\cite{Leturcq2009} Phonons also give rise to spin relaxation within a time $T_1$, which is important for spintronics devices \cite{Trauzettel2007,Khaetskii2001,Kuemmeth2008,Struck2010}.
The spin-orbit interaction admixes different spin states ($\uparrow,\downarrow$) and electron orbits $k$ [see Eq. (\ref{SO}) below].
%, such that electron-phonon coupling $H_{\text{EPC}}$ can mediate the Zeeman energy
%The effective spin-phonon interaction $H_{\text{s-p}}$ Then, electron-phonon coupling $H_{\text{EPC}}$ couples the Zeeman-split spin states ($\uparrow,\downarrow$) to the phonon bath with phonon numbers $n_{\omega}$, into which the Zeeman energy $g\mu_BB=\hbar\omega$, where $g$ is the electron g-factor in graphene and $\mu_B$ denotes Bohr's magneton, is released. The rate for spin relaxation via emission of a phonon with energy $\hbar\omega$ is given by
As a consequence, the electron-phonon coupling $H_{\text{EPC}}$ can mediate the Zeeman energy $g\mu_BB=\hbar\omega$, where $g$ is the electron $g$-factor in graphene and $\mu_B$ denotes Bohr's magneton, to the phonon bath with phonon numbers $n_{\omega}$. The rate for spin relaxation via emission of a phonon with energy $\hbar\omega$ is given by
\begin{eqnarray}
\frac{1}{T_1}=\frac{2\pi}{\hbar}|\langle k\!\downarrow,n_{\omega}+1|H_{\text{EPC}}|k\!\uparrow,n_{\omega}\rangle|^2\rho_{\text{states}}(\hbar\omega)\, ,\label{eq0}
\end{eqnarray}
with an explicit dependence on the phonon density of states $\rho_{\text{states}}$.
Several mechanisms contribute to $T_1^{-1}$ and in some cases the deformation potential\cite{Struck2010,Mariani2009} gives the dominant contribution. If the Zeeman energy lies within the energy gap of GNR phonons with fixed boundaries and if the temperature is sufficiently low, the spin lifetime obtained from Eq. (\ref{eq0}) diverges due to a vanishing density of states.
%
%As an example for $H_{\text{EPC}}$, we present the deformation potential in graphene nanoribbons and  its influence on $T_1$ for small magnetic fields.
%

\section{Continuum model in 2D}
Low-energy acoustic phonons at the center of the Brillouin zone have a wavelength much larger than atomic distances and thus can be derived from continuum mechanics. The carbon atoms in graphene lie within a two-dimensional surface and this property is conserved upon deformations, making graphene a quasi-two-dimensional material in three-dimensional (3D) real space. Consequently, all components of the displacement field $\vec{u}(\vec{r})%=(u_x,u_y,u_z)
$ can be nonzero but the components $u_{iz}$ of the strain tensor $u_{ik}=(\partial_iu_k+\partial_ku_i)/2$ vanish identically. While $u_{xz}$ and $u_{yz}$ are known to vanish for thin plates in the $xy$ plane in general, the monatomic thickness of graphene implies that $u_{zz}$ must vanish as well.
%
%The disappearance of $u_{zz}$ becomes obvious when considering the relative volume change upon deformation, given by the trace $u_{ii}$: for a 2D lattice, there is no lowest order contribution from neighboring atoms shifting out-of-plane.
%
With $u_{iz}\equiv0$, the elastic Lagrangian density of monolayer graphene is given by
\begin{eqnarray}
{\cal L} = {\cal T} -{\cal V} =\frac{\rho}{2}\dot{\bf u}^2 - \frac{\kappa}{2} \left(\Delta u_z\right)^2-\frac{\lambda}{2} u_{ii}^2 -\mu u_{ik}^2 , \label{eq1}
\end{eqnarray}
where $\Delta=\partial_x^2+\partial_y^2$, the sum convention with $u_{ii}=u_{xx}+u_{yy}+u_{zz}$ and $u_{ik}^2=u_{xx}^2 + u_{xy}^2 + \cdots$ has been used, $\rho$ is the surface mass density, and $\kappa$ is the bending rigidity.\cite{LandauLifschitz7.7,Suzuura2002,Mariani2008,Mariani2009} Note that the 3D bulk elastic constants have been replaced by their 2D analogs $\lambda=2h\mu_{\text{3D}}\lambda_{\text{3D}}/(2\mu_{\text{3D}}+\lambda_{\text{3D}})$ and $\mu=h \mu_{\text{3D}}$ where $h$ is the plate thickness. The bulk and shear moduli are then given as $B=\lambda+\mu$ and $\mu$, respectively.

Application of the Euler-Lagrange formalism to the functional (\ref{eq1}) leads to the coupled set of differential equations for in-plane modes
\begin{eqnarray}
\begin{split}
\rho\,\ddot{u}_x=(B+\mu)\partial_x^2u_x+\mu\,\partial_y^2u_x+B\,\partial_x\partial_yu_y\,,\\
\rho\,\ddot{u}_y=(B+\mu)\partial_y^2u_y+\mu\,\partial_x^2u_y+B\,\partial_x\partial_yu_x\,,
\end{split}
\label{eq2}
\end{eqnarray}
which are decoupled from the differential equation for the out-of-plane modes,
\begin{eqnarray}
\rho\,\ddot{u}_z=-\kappa\left(\partial_x^2+\partial_y^2\right)^2u_z\,.\label{eq3}
\end{eqnarray}
Assuming nanoribbon alignment with the $y$ axis, fixed boundaries are described by
\begin{eqnarray}
u_x=u_y=0&&\quad\text{(in plane),}\\
u_z=\partial_xu_z=0&&\quad\text{(out of plane)}
\end{eqnarray}
%
%$u_x=u_y=0$ (in-plane) and $u_z=\partial_xu_z=0$ (out-of-plane) 
at $x=\pm W/2$; see Fig.~\ref{pic1} (a). While these boundary conditions hold for both 2D and 3D lattices, we emphasize that lattice dimensionality does affect free boundaries. For free edges in 2D it is required that,
at $x=\pm W/2$,
\begin{eqnarray}
\left.
\begin{array}{r}
\partial_xu_x+\sigma\partial_yu_y=0\\
\partial_xu_y+\partial_yu_x=0
\end{array}
\right\}&&\text{(in plane),}\\
\left.
\begin{array}{r}
\partial_x^3u_z+(2-\sigma)\partial_x\partial_y^2u_z=0\\
\partial_x^2u_z+\sigma\partial_y^2u_z=0
\end{array}
\right\}&&\text{(out of plane)},
\end{eqnarray}
%
%$\partial_xu_x+\sigma\partial_yu_y=\partial_xu_y+\partial_yu_x=0$ (in-plane) and $\partial_x^3u_z+(2-\sigma)\partial_x\partial_y^2u_z=\partial_x^2u_z+\sigma\partial_y^2u_z=0$ (out-of-plane), 
where the quantity $\sigma$ denotes Poisson's ratio, Fig. \ref{pic1} (b). Together with Young's modulus $\mathpzc{E}=h\mathpzc{E}_{\text{3D}}$, $\sigma$ relates to the bulk and shear moduli as
\begin{eqnarray}
B=\frac{\mathpzc{E}}{2(1-\sigma)}\,,\quad\mu=\frac{\mathpzc{E}}{2(1+\sigma)}\,.\label{con}
\end{eqnarray}
\begin{figure}[t]\centering\includegraphics[width = 0.48\textwidth]{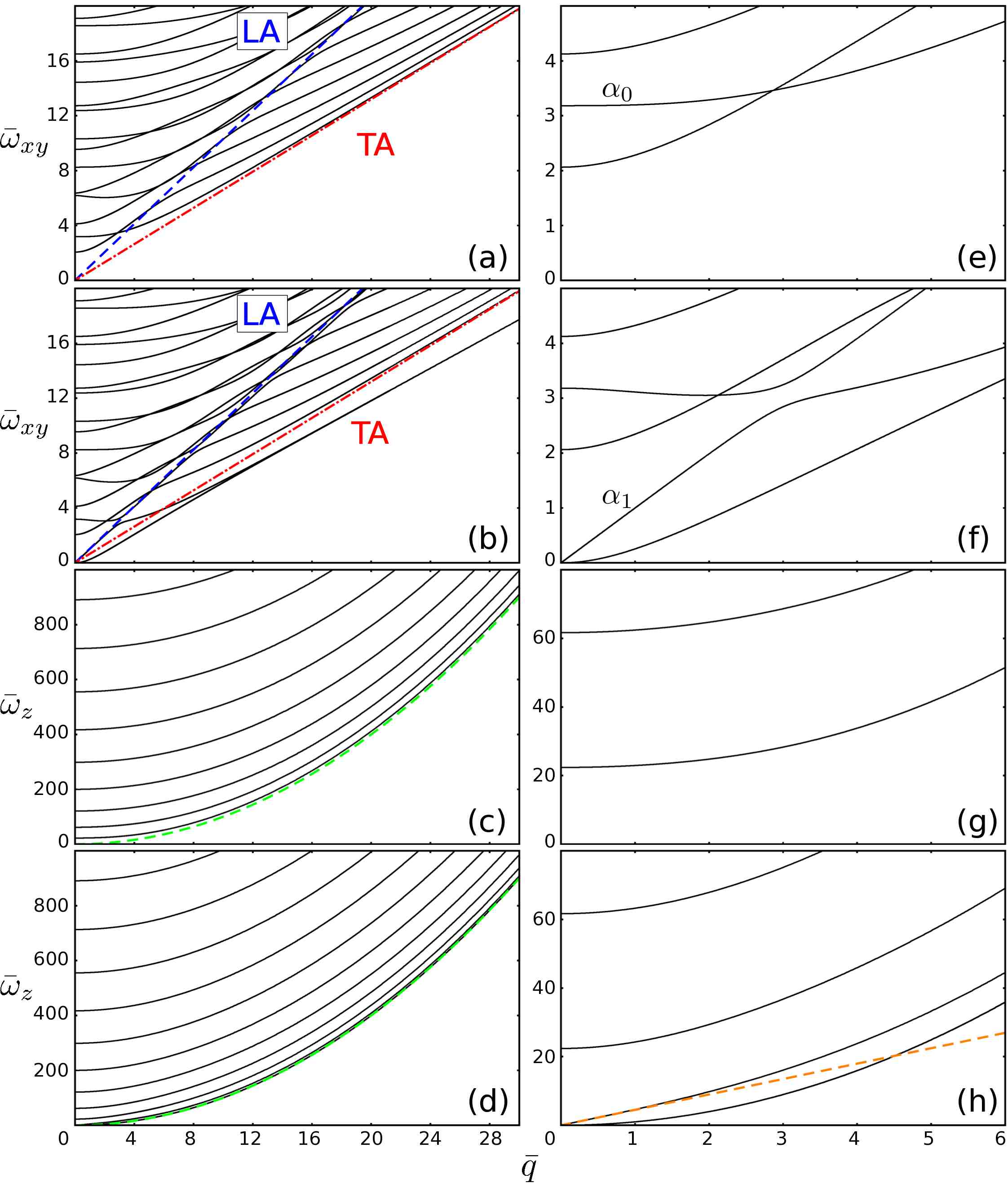}\caption{(Color online) Dispersion relations obtained from the procedure described in Sec. \ref{class}. The wavenumber $q$ is given by $\bar{q}=qW$ and the frequency $\omega$ of in-plane (out-of-plane) phonons by $\bar{\omega}_{xy}=\omega\sqrt{\rho/\mathpzc{E}}\,W$ ($\bar{\omega}_z=\omega\sqrt{\rho/\kappa}\,W^2$). (a) In-plane modes with fixed boundaries. (b) In-plane modes with free boundaries. (c) Out-of-plane modes with fixed boundaries. (d) Out-of-plane modes with free boundaries. (e)-(h) Dispersion relations (a)-(d) at the center of the Brillouin zone. (a),(c),(e),(g) Modes with fixed boundaries exhibit a gap. (b),(d),(f),(h) Modes with free boundaries are gapless. (a),(b) Despite the coupling of transverse and longitudinal modes, we find predominantly longitudinal and transverse modes on lines which we label LA (dashed blue line) and TA (dash-dotted red line), respectively. (c),(d) Independent of the boundaries, out-of-plane modes disperse quadratically for large wave numbers (dashed green line). (h) Free out-of-plane modes feature a branch with linear dispersion at the zone center (dashed orange line).
}\label{pic3}\end{figure}

\section{Classical solution}\label{class}
Typically, the length of a graphene nanoribbon exceeds its width many times,\cite{Jiao2009,Kosynkin2009,Shivaraman2009,Li2008} $L\gg W$, thus allowing for a plane wave ansatz along the $y$ direction with periodic boundaries. Due to their decoupling, in-plane modes $u_{x/y}(x,y,t)=f_{x/y}(x)\text{exp}[i(qy-\omega t)]$ and out-of-plane modes $u_z(x,y,t)=f_z(x)\text{exp}[i(qy-\omega t)]$ can be treated separately.\cite{footnote}

Exploiting the plane wave ansatz and denoting the $i$-th derivative of $f$ as $f^{(i)}$, Eq. (\ref{eq2}) can be written as $\mathcal{M}_{xy}(f_x,f_y)=-\rho\omega^2(f_x,f_y)$, where
\begin{eqnarray}
\!\!\mathcal{M}_{xy}\!:\!\begin{pmatrix}f_x\\f_y\end{pmatrix}\!\mapsto\!\begin{pmatrix}(B+\mu)f_x^{(2)}-\mu q^2f_x+iBqf_y^{(1)}\\-(B+\mu)q^2f_y+\mu f_y^{(2)}+iBqf_x^{(1)}\end{pmatrix}\!.\label{eq5}
\end{eqnarray}
The general solution of this eigenvalue problem is $(f_x,f_y)=\sum_{i=1}^4c_i\vec{a}_i\text{exp}[\lambda_ix]$, with $\vec{a}_1=(1,iq/\lambda_1)$, $\vec{a}_2=(1,iq/\lambda_2)$, $\vec{a}_3=(1,i\lambda_3/q)$, $\vec{a}_4=(1,i\lambda_4/q)$, and $\lambda_{1,2}=\pm\sqrt{q^2-\rho\omega^2/(B+\mu)}$, $\lambda_{3,4}=\pm\sqrt{q^2-\rho\omega^2/\mu}$.

\begin{figure}[t]\centering\includegraphics[width = 0.48\textwidth]{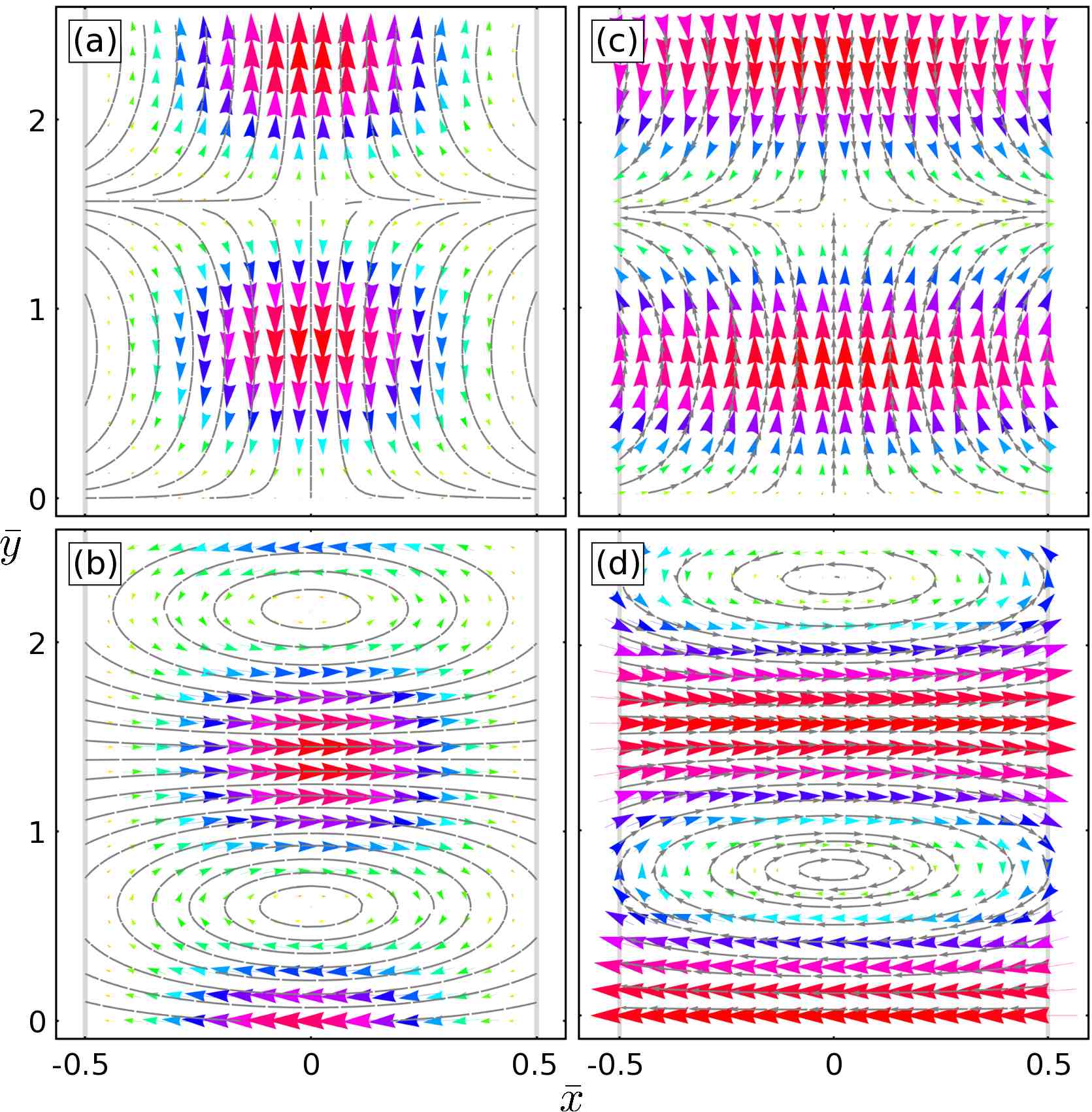}\caption{(Color online) Displacement vector field $(u_x,u_y)$ of in-plane modes. Size and color of the arrows indicate the magnitude of the local deformation. We use the dimensionless coordinates $\bar{x}=x/W$ and $\bar{y}=y/W$. (a),(b) Fixed boundaries. (c),(d) Free boundaries. (a),(c) Predominantly longitudinal modes. (b),(d) Predominantly transverse modes.}\label{pic4}\end{figure}

Fixed boundaries are characterized by $f_x(\pm W/2)=f_y(\pm W/2)=0$ and by virtue of the $\lambda_i$, the set of linear equations deriving from these boundary conditions depends on the parameters $q$ and $\omega$. A numerical treatment of this linear system yields the dispersion relation [Figs. \ref{pic3}(a) and \ref{pic3}(e)] as well as the coefficients $c_i$ for the explicit form of the in-plane mode with fixed boundaries [Figs. \ref{pic4}(a) and \ref{pic4}(b)]. Other boundary conditions and the out-of-plane modes can be treated likewise. The eigenvalue problem obtained from (\ref{eq3}) is $\mathcal{M}_zf_z=(\rho\omega^2/\kappa-q^4)f_z$, where the map $\mathcal{M}_z$ and its eigenfunctions and eigenvalues are given by
\begin{eqnarray}
%&&
\mathcal{M}_z:\,f_z\mapsto f_z^{(4)}-2q^2f_z^{(2)}%\,,
\label{eq4}
%\\
%&&f_z=\sum_{i=1}^4d_ie^{\lambda_ix}\,,\quad\lambda_i=\pm\sqrt{q^2\pm\omega\sqrt{\rho/\kappa}}\,.
\end{eqnarray}
and $f_z=\sum_{i=1}^4d_ie^{\lambda_ix}$ with $\lambda_i=\pm\sqrt{q^2\pm\omega\sqrt{\rho/\kappa}}$.

\section{Mode orthonormality and quantization}
In order to quantize the vibrational spectrum of the graphene nanoribbon in terms of phonon creation and annihilation operators, the eigenfunctions of the original differential operators [Eqs. (\ref{eq2}) and (\ref{eq3})] must be orthogonal. While orthogonality of eigenmodes with different wavenumbers $q$ follows from the plane wave ansatz, eigenmodes with same $q$ require orthogonal functions $(f_{(\alpha,q),x},\,f_{(\alpha,q),y})$ and $f_{(\alpha,q),z}$. The index ${(\alpha,q)}$ labels the phonon branch $\alpha$ and the wavenumber $q$ of a specific eigenmode.

The map (\ref{eq5}) is Hermitian and hence has orthogonal eigenfunctions if and only if
the scalar product
% $\langle(f_x,f_y),\mathcal{M}_{xy}(f_x,f_y)\rangle$ is real
%
%\begin{eqnarray}
$\int_{-W/2}^{+W/2}\text{d}{x}\left(f^*_x,f^*_y\right)\mathcal{M}_{xy}(f_x,f_y)^{\text{T}}$
%\in\mathbb{R}
%\end{eqnarray}
is real
for all vector functions $(f_x,f_y)$ in the domain of $\mathcal{M}_{xy}$. One easily shows via partial integration that $\mathcal{M}_{xy}$ is Hermitian if and only if the boundary terms satisfy
\begin{eqnarray}
\left.(B+\mu)f^*_xf_x^{(1)}+\mu f^*_yf_y^{(1)}+iBqf^*_xf_y\right|_{-\frac{W}{2}}^{+\frac{W}{2}}\in\mathbb{R}
\end{eqnarray}
and that both fixed and free boundaries do indeed satisfy this condition.

The general in-plane displacement field is
\begin{eqnarray}
\vec{u}_{\parallel}=\sum_{\alpha,q}r_{(\alpha,q)}\left(f_{(\alpha,q),x}\vec{e}_x+f_{(\alpha,q),y}\vec{e}_y\right)e^{iqy}\,,\label{eq6}
\end{eqnarray}
where the harmonic time dependence has been absorbed in the normal coordinate. Using the orthogonality relations mentioned above, one can resolve the normal coordinate and derive the Lagrangian and the canonical momentum. The identification
\begin{eqnarray}
r_{(\alpha,q)}=\sqrt{\hbar/2\rho LW\omega_{(\alpha,q)}}(b_{(\alpha,q)}+b^{\dagger}_{(\alpha,-q)})\,,\label{eq7}
\end{eqnarray}
where $b^{\dagger}_{(\alpha,q)}$ ($b_{(\alpha,q)}$) creates (annihilates) an $(\alpha,q)$-phonon, complies with coordinate-momentum commutation relations, and allows for a quantum mechanical formulation of (\ref{eq6}). Quantization of the out-of-plane modes is achieved in the very same way. The Hermiticity of $\mathcal{M}_z$ follows from 
\begin{eqnarray}
\left.f^*_zf_z^{(3)}-f_z^{*(1)}f_z^{(2)}-2q^2f^*_zf_z^{(1)}\right|_{-\frac{W}{2}}^{+\frac{W}{2}}\in\mathbb{R}
\end{eqnarray}
and, as above, fixed as well as free boundaries do satisfy this condition. The general out-of-plane displacement is given by
\begin{eqnarray}
\vec{u}_{\bot}=\sum_{\alpha,q}r_{(\alpha,q)}f_{(\alpha,q),z}\vec{e}_ze^{iqy}\,.
\end{eqnarray}
%A detailed discussion of the phonon spectra will follow in Sec. \ref{discuss}.

\section{Discussion of phonon spectra}\label{discuss}
As specific values for sound velocities, etc., depend on the elastic constants, we shall first discuss these constants before turning to the properties of acoustic phonons. Due to their decoupling, in-plane and out-of-plane phonons can be treated separately. For each case we will consider fixed and free boundaries.

\subsection{Elastic constants}
For graphene, most elastic constants remain to be settled by experiment and some seem to exhibit a temperature dependence, which we do not take into account here. Moreover, a consistent set of constants must respect Eq. (\ref{con}).
%relations between these constants need to be respected.
The Zeeman energy for typical laboratory magnetic fields ($\sim1\,\text{T}$) marks $\sim1\,\text{K}$ as the temperature range where the phonon properties can be probed via electron spin relaxation.

Cited values for Poisson's ratio $\sigma$ of graphene range from\cite{Reddy2006} 0.145 to 0.416 but accumulate around $\sigma=0.16$, which we use in our calculations.\cite{Lee2008,Faccio2009,Kudin2001} Young's modulus of a quasi-two-dimensional material, $\mathpzc{E}=\mathpzc{E}_{\text{3D}}h$, follows from its corresponding 3D bulk value and its associated thickness $h$. While the most common literature value\cite{Lee2008,Faccio2009,Kudin2001} of $\mathpzc{E}_{\text{3D}}$ for graphene is $1\,\text{TPa}$, a much smaller value, $0.5\,\text{TPa}$, has been found in at least one experiment.\cite{Frank2007} We use $\mathpzc{E}=3.4\,\text{TPa\AA}$, the product of $1\,\text{TPa}$ and the interlayer spacing of graphite, $3.4\,\text{\AA}$. Substituting our choices for $\sigma$ and $\mathpzc{E}$ into Eq. (\ref{con}), we find $B=12.6\,\text{eV/\AA}\! ^2$ and $\mu=9.1\,\text{eV/\AA}\!^2$ for the bulk and shear moduli, respectively, in agreement with literature values.\cite{Gazit2009,Kudin2001} All these values are in agreement with results of simulations for zero temperature.\cite{Zakharchenko2011}
%According to Ref. [\onlinecite{Faccio2009}], $\mathpzc{E}$ increases for GNR widths less than $20\,\text{\AA}$. Here, we are interested in GNRs with typically $W\sim30\,\text{nm}$ such that this effect need not be considered.

The bending rigidity of graphene, $\kappa$, is mainly determined by the out-of-plane $p_z$ orbitals such that it cannot be inferred from other elastic constants. It has been shown that $\kappa$ decreases with increasing temperature.\cite{Liu2009} Literature values for zero temperature\cite{Fasolino2007,Gazit2009,Liu2009,Kudin2001} range from $0.85$ to $1.22\,\text{eV}$ and we choose $\kappa=1.1\,\text{eV}$.

The mass density of graphene, $\rho=7.61\times10^{-7}\, {\rm kg}/{\rm m}^2$, follows directly from the atomic weight of natural carbon, $12.01\,\text{u}$, and the interatomic distance in graphene, $1.42\,\text{\AA}$.

\subsection{In-plane phonons}
The dispersion relation of in-plane modes with fixed boundaries is gapped and features infinitely many branches with different energies originating from the zone center, Figs. \ref{pic3}(a) and \ref{pic3}(e). The gap relates to the energy necessary for fixing the boundaries and is given by $2.1\,\hbar\sqrt{\mathpzc{E}/\rho}\,/W$. For $W=30\,{\rm nm}$, this gap will be $1.0\,{\rm meV}$, corresponding to a magnetic field of $8.4\,\text{T}$. For large wave numbers, all branches converge to a common line, which we label TA. A second line, labeled LA, is supported by different branches throughout the dispersion relation. Due to coupling at the ribbon boundaries there are no purely transverse or longitudinal modes. However, we do find that the modes on the TA (LA) line have predominantly transverse (longitudinal) character, Figs. \ref{pic4}(a) and \ref{pic4}(b). The corresponding sound velocities are $v_{\text{LA}}=22\,{\rm km}/{\rm s}$ and $v_{\text{TA}}=14\,{\rm km}/{\rm s}$, independent of the ribbon width. These values and the ratio $v_{\text{LA}}/v_{\text{TA}}=1.6$ are in good agreement with previous calculations for bulk graphene\cite{Falkovsky2008} ($19.5\,{\rm km}/{\rm s}$, $12.2\,{\rm km}/{\rm s}$) and carbon nanotubes\cite{Suzuura2002} ($19.9\,{\rm km}/{\rm s}$, $12.3\,{\rm km}/{\rm s}$). % For the ratio $v_{\text{LA}}/v_{\text{TA}}=1.6$, we even find very good agreement.
Nevertheless, we point out that our sound velocities are proportional to $\sqrt{\mathpzc{E}/\rho}$, a value that is still under discussion for graphene. The approach to linear, bulk-like behavior is expected for large wave number, where the finite ribbon width appears like bulk for short-wavelength phonons. 

For free boundaries, the dispersion relation of in-plane modes is ungapped and the two branches that start at zero energy converge slightly below the TA line, Figs. \ref{pic3}(b) and \ref{pic3}(f). The sound velocities and linear behavior for large wave number do not depend on boundary conditions, as one would expect from the same argument as above. Predominantly transverse and predominantly longitudinal modes are shown in Figs. \ref{pic4}(c) and \ref{pic4}(d). The typical zero-point motion amplitude of in-plane modes is $40\,{\rm fm}$.

\subsection{Out-of-plane phonons}
The dispersion relation of out-of-plane modes with fixed boundaries is shown in Figs. \ref{pic3}(c) and \ref{pic3}(g). The gap due to the fixed boundary conditions is given by $22.4\,\hbar\sqrt{\kappa/\rho}\,/W^2$, which yields $7.9\,\mu{\rm eV}$ for $W=30\,{\rm nm}$. The corresponding magnetic field is $68\,\text{mT}$. There are infinitely many branches that correspond to different transverse excitations, Figs. \ref{pic5}(a) and \ref{pic5}(b). Again, away from the zone center, all branches approach bulk behavior, that is, a quadratic dispersion for out-of-plane modes.\cite{Falkovsky2008}

Similarly, the out-of-plane modes with free boundaries disperse quadratically as in the bulk, for large wave numbers,  Figs. \ref{pic3}(d) and \ref{pic3}(h). The dispersion relation is gapless and one branch exhibits a finite sound velocity at the zone center. This sound velocity amounts to about $70\,{\rm m}/{\rm s}$ for $W=30\,{\rm nm}$, is proportional to $\sqrt{\kappa/\rho}\,/W$, and hence goes to zero for large $W$, again in agreement with bulk graphene. The typical zero-point motion amplitude of out-of-plane modes is $0.4\,{\rm pm}$.

\section{Deformation potential and spin relaxation}\label{relax}
Several mechanisms contribute to spin relaxation: out-of-plane modes via direct spin-phonon coupling and in-plane phonons via the deformation potential and bond-length change\cite{Struck2010}. Due to inversion symmetry, piezoelectric coupling does not occur in graphene. Here, we discuss the deformation potential, which gives the dominant contribution to $T_1^{-1}$ under certain conditions.
%In leading order, it only depends on in-plane phonons,
%\begin{eqnarray}
%H_{\text{EPC}}=g_\text{D}\vec{\nabla}\cdot\vec{u}_{\parallel}(x,y)\,,\label{eq8}
%\end{eqnarray}
%where $g_D\approx30\,{\rm eV}$ is the coupling strength\cite{Struck2010,Suzuura2002} and $\vec{\nabla}=(\partial_x,\partial_y)$.

%While coupling to out-of-plane phonons is mediated by the intrinsic spin-orbit interaction, in-plane phonons couple to the spin via the Rashba-type spin-orbit mechanism. This allows to augment the coupling strength to in-plane phonons via an external gate electric field.

%\subsection{Spin relaxation via the deformation potential}
We find that any given in-plane phonon branch $\alpha$ couples either via bond-length change or via the deformation potential, depending on whether its displacement field is even or odd in the $x$ coordinate. The branch that originates from $\bar{\omega}_{xy}=3.2$ in Fig. \ref{pic3} (e), labeled $\alpha_0$, has a flat dispersion at the zone center and couples to the spin only via the deformation potential. As a consequence of the density of states in Eq. (\ref{eq0}), this mechanism will give the dominant contribution if the magnetic field is tuned to a value where the Zeeman energy is close to the Van Hove singularity of $\alpha_0$ and coupling to out-of-plane modes is weak. Van Hove singularities also occur for out-of-plane modes at different values of $\bar{\omega}_z$, Figs. \ref{pic3}(c) and \ref{pic3}(g). However, $\bar{\omega}_{xy}$ and $\bar{\omega}_z$ scale differently with $W$, which allows us to choose a ribbon width where there is a singularity for in-plane modes ($\alpha_0$) but not for out-of-plane modes. This situation will be discussed below.

For the branch labeled $\alpha_1$ in Fig. \ref{pic3} (f), which is linear near the zone center, the spin couples to phonons only via the deformation potential, as well. Even though its density of states is finite, we discuss its contribution to Eq. (\ref{eq0}) as it is in accordance with previous results for semiconductor quantum dots.\cite{Khaetskii2001}
%For $\alpha_0$, this condition is fulfilled at $\bar{\omega}_{xy}\gtrsim3.2$ which relates to a magnetic field of about $12.8\,\text{T}$ for $W=30\,\text{nm}$ and $3.8\,\text{T}$ for $W=100\,\text{nm}$ and hence can be accessed experimentally.

In leading order, the deformation potential depends only on in-plane phonons,
\begin{eqnarray}
H_{\text{EPC}}=g_\text{D}\vec{\nabla}\cdot\vec{u}_{\parallel}(x,y)\,,\label{eq8}
\end{eqnarray}
where $g_D\approx30\,{\rm eV}$ is the coupling strength\cite{Struck2010,Suzuura2002} and $\vec{\nabla}=(\partial_x,\partial_y)$. The deformation potential is independent of the electron spin ($\uparrow,\downarrow$) but it does couple different electron orbits ($k$). As a consequence, $H_{\text{EPC}}$ couples to spin indirectly when Rashba-type spin-orbit interaction, $H_{\text{SO}}$, is taken into account. In lowest order, the spin-orbit-perturbed electronic states are given by
\begin{eqnarray}
|k\!\uparrow\rangle=|k\!\uparrow\rangle^{\text{\tiny$(0)$}}+\sum_{k'\neq k}|k'\!\downarrow\rangle^{\text{\tiny$(0)$}}\frac{^{\text{\tiny$(0)$}}\langle k'\!\!\downarrow\!|H_{\text{SO}}|k\!\uparrow\rangle^{\text{\tiny$(0)$}}}{E_k-E_{k'}+g\mu_BB}\,,\label{SO}
\end{eqnarray}
where the superscript $(0)$ indicates unperturbed product states. Using these spin-orbit admixed states, we find
\begin{eqnarray}
\langle k\!\downarrow\!|H_{\text{EPC}}|k\!\uparrow\rangle\hspace{+6.0cm}\label{EPC}\\
=\sum_{k'\neq k}\left[\frac{(H_{\text{EPC}})_{kk'}(H_{\text{SO}})_{k'k}^{\downarrow\uparrow}}{E_k-E_{k'}+g\mu_BB}+\frac{(H_{\text{EPC}})_{k'k}(H_{\text{SO}})_{kk'}^{\downarrow\uparrow}}{E_k-E_{k'}-g\mu_BB}\right]\!,\nonumber
\end{eqnarray}
where we denote the numerator in Eq. (\ref{SO}) as $(H_{\text{SO}})_{k'k}^{\downarrow\uparrow}$ and the spin-conserving transitions of $H_{\text{EPC}}$ accordingly. This is the matrix element required to calculate the relaxation rate in Eq. (\ref{eq0}).

We find that for a given $k'$ the two terms in Eq. (\ref{EPC}) exactly cancel each other at $B=0$. This effect is known as Van Vleck cancellation and is expected for time-reversal-symmetric systems. Moreover, $(H_{\text{SO}})_{k'k}^{\downarrow\uparrow}$ vanishes if both $k$ and $k'$ are even or odd at the same time.

%GNRs with free edges have ungapped phonon spectra. Due to energy conservation, the Zeeman energy must match the phonon energy, such that only the two lowest branches in Fig. \ref{pic3} (f) are accessible for low magnetic fields ($B\lesssim100\,\text{mT}$). The branch $\alpha_1$, which is linear at the center, gives the sole contribution since the other branch is a pure shear mode. Due to its linear dispersion, we find $B\propto\omega\propto q$ and a constant density of states on this branch. In this case, the matrix element in Eq. (\ref{EPC}) scales as $B^{2.5}$: one order in $B$ arising from Van Vleck cancellation, dipole approximation, and the gradient in Eq. (\ref{eq8}) each\cite{VanVleck1940,Trauzettel2007,Struck2010,Droth2010}, reduced by the prefactor $\omega^{-0.5}\propto B^{-0.5}$ in Eq. (\ref{eq7}). Consequently, for low magnetic fields, the contribution of deformation potential and spin-orbit coupling to the spin relaxation rate (\ref{eq0}) scales with $B^5$.
%However, this effect might be screened by contributions that arise from bond-length change and direct spin-phonon coupling.

%\subsection{Spin lifetime for different situations}
For fixed GNR edges, the phonon spectrum is gapped. In the range $3.2\leq\bar{\omega}_{xy}\leq3.3$, the branch $\alpha_0$ shows an almost flat dispersion. Its sound velocity increases as $v_{\alpha_0}\propto q^2$ such that the corresponding density of states behaves as $(\rho_{\text{states}})_{\alpha_0}\propto q^{-2}$. The matrix element (\ref{EPC}) varies as $q\,B^{0.5}$: the dipole approximation gives rise to one order in $q$ and Van Vleck cancellation\cite{VanVleck1940,Trauzettel2007,Struck2010,Droth2010} to one order in $B$, reduced by $\omega^{-0.5}\propto B^{-0.5}$ due to the prefactor in Eq. (\ref{eq7}). In total, the contribution to the spin relaxation rate (\ref{eq0}) is proportional to the magnetic field. Due to the Van Hove singularity of $\alpha_0$ at $\bar{\omega}_{xy}=3.2$, we expect that $T_1^{-1}\propto B$ is the dominant behavior in the range $3.8\leq B\leq4.0\,\text{T}$ ($12.8\leq B\leq13.2\,\text{T}$) for $W=100\,\text{nm}$ ($W=30\,\text{nm}$), where the density of states of out-of-plane modes is relatively small. These are accessible laboratory magnetic fields and hence allow for experimental examination of our results.

If the magnetic field is tuned to a value where the Zeeman energy lies within the gap of both in-plane and out-of-plane phonons [Figs. \ref{pic3}(e) and \ref{pic3}(g)], the electron spin cannot flip due to phonon emission.
Then, multiple-phonon processes, where the Zeeman energy corresponds to the difference between an absorbed and an emitted phonon, become important.
Again due to the gap, these processes can be frozen out if the temperature $T$ is low enough. As discussed in Sec. \ref{discuss}, the very soft out-of-plane modes have a much smaller gap, which therefore imposes a tighter condition and which scales as $W^{-2}$. Assuming $W=30\,\text{nm}$, the spin lifetime inferred from Eq. (\ref{eq0}) diverges for $B<68\,\text{mT}$ and $T\ll90\,\text{mK}$. Very narrow GNRs with $W=10\,\text{nm}$ are studied experimentally, as well.\cite{Wang2011} Accordingly, the requirements for such a ribbon would be $B<0.61\,\text{T}$ and $T\ll0.8\,\text{K}$.

GNRs with free edges have ungapped phonon spectra. Due to energy conservation, the Zeeman energy must match the phonon energy, such that only the two lowest branches in Fig. \ref{pic3} (f) are accessible for low magnetic fields ($B\lesssim100\,\text{mT}$). The branch $\alpha_1$ couples only via the deformation potential and the other branch is a pure shear mode. Due to its linear dispersion, we find $B\propto\omega\propto q$ and a constant density of states for $\alpha_1$. The matrix element in Eq. (\ref{EPC}) scales as $B^{2.5}$: one order in $B$ arising from each of the Van Vleck cancellation, dipole approximation, and the gradient in Eq. (\ref{eq8}),\cite{footnote2} again reduced by the prefactor $\omega^{-0.5}\propto B^{-0.5}$ in Eq. (\ref{eq7}). Consequently, for low magnetic fields, the contribution of deformation potential and spin-orbit coupling to the spin relaxation rate (\ref{eq0}) scales with $B^5$. In semiconductors, $T_1^{-1}\propto B^5$ holds, as well.\cite{Khaetskii2001}%\\[+1cm]
%However, this effect might be screened by contributions that arise from bond-length change and direct spin-phonon coupling.

\section{Conclusion}
Acoustic phonons are relevant for many GNR applications and can be probed with established techniques.\cite{Oshima1988,Mohr2007} Using a continuum model that accounts for the monatomic thickness of graphene, we derive boundary conditions that lead to Hermitian wave equations. We focus on two types of boundary configurations: fixed and free boundaries. We explicitly give the corresponding classical solutions and, ensuring Hermiticity, infer a quantum theory with ribbon phonon creation and annihilation operators. Free boundaries lead to ungapped dispersion relations. In contrast, fixed boundaries lead to a gapped phonon dispersion of both in-plane and out-of-plane modes, which is most suitable for achieving high mobilities as well as long spin lifetimes. Regardless of the boundary configuration, all dispersion relations approach bulk behavior for wavelengths small compared to the ribbon width. Sound velocities that relate to transverse and longitudinal acoustical in-plane ribbon modes are in good accordance with values for bulk graphene.
%
%
%As an example for electron-phonon coupling, which is relevant for spintronics and transport applications, we present the deformation potential in graphene nanoribbons.
%
%For magnetic fields up to $B\sim100\text{ mT}$, its contribution to the spin relaxation rate scales with $B^5$.
%
We also study phonon-induced spin relaxation in GNRs. We find that, if the Zeeman energy is tuned close to a Van Hove singularity of the density of states of in-plane phonons, the deformation potential can be the dominant effect for spin relaxation. In this case, it should be possible to probe our predicted behavior for $T_1$ experimentally. If the Zeeman energy lies within the gap of both in-plane and out-of-plane phonons with fixed boundaries and for low enough temperatures, coupling to the lattice is inhibited such that the spin lifetime obtained form Eq. (\ref{eq0}) diverges.
%For free boundaries and low magnetic fields ($B\lesssim100\text{ mT}$), we find that the electron-phonon coupling mechanism with the strongest coupling strength, the deformation potential, gives rise to a $B^5$ behavior of the spin relaxation rate.

\section{acknowledgements}
We thank András Pályi and Michael Pokojovy for helpful discussions and acknowledge funding from the ESF within the EuroGRAPHENE project CONGRAN.


\begin{thebibliography}{}
\bibitem{footnote}
The physical displacement (see Figs. \ref{pic5} and \ref{pic4}) is obtained by taking the real part.
\end{thebibliography}
\end{document}